\documentclass[aps,pxl,twocolumn,superscriptaddress,longbibliography]{revtex4-1}

 \newcommand{\bra}[1]{\left< #1\right|}

\usepackage{hyperref}
\usepackage{graphicx}
\usepackage{amsmath}
\usepackage{xcolor}
\usepackage{bm}
\usepackage{color}
\usepackage{enumitem} 
\usepackage[T1]{fontenc}
\usepackage{lipsum}
\usepackage{amsmath, array} 
\usepackage[normalem]{ulem}
\usepackage{soul,xcolor}
\usepackage{lineno}
\usepackage{physics}

\usepackage{color}
\usepackage{soul,xcolor}
\usepackage[normalem]{ulem}
\definecolor{black}{rgb}{0,0,0}
\definecolor{blue}{rgb}{0,0,1}
\definecolor{green}{rgb}{0,1,0}
\definecolor{red}{rgb}{1,0,0}
\definecolor{brown}{rgb}{0.4,0.2,0}
\definecolor{darkgreen}{rgb}{0,0.7,0}
\definecolor{darkblue}{rgb}{0.0,0.0,0.5}
\definecolor{red}{rgb}{1,0,0}
\definecolor{deepmagenta}{rgb}{0.8, 0.0, 0.8}

\newcommand{\hidetext}[1]{{{\color{blue}\sout{}}}}
\newcommand{\veck}{\mathbf k}

\newcommand{\vecr}{\mathbf r}

\newcommand{\bea}{\begin{eqnarray}}
\newcommand{\ea}{\end{eqnarray}}
\newcommand{\eea}{\end{eqnarray}}

\usepackage{amssymb}

\newcommand{\sumint}[1]

\date{\today}
\begin{document}
\title{Rydberg impurity in a Fermi gas: Quantum statistics and rotational blockade}

\author{John Sous}\thanks{Current address: Department of Physics, Columbia University, New York, New York 10027, USA.
\\Email: js5530@columbia.edu.
} 
\affiliation{ITAMP, Harvard-Smithsonian Center for Astrophysics, Cambridge, Massachusetts 02138, USA} 
 \affiliation{Department of Physics and Astronomy, University of British Columbia, Vancouver, British Columbia V6T 1Z1, Canada}

\author{H. R. Sadeghpour} 
\affiliation{ITAMP, Harvard-Smithsonian Center for Astrophysics, Cambridge, Massachusetts 02138, USA}

\author{T. C. Killian} \affiliation{Department of Physics \& Astronomy and Rice Center for Quantum Materials, Rice University, Houston, TX 77251, USA}

\author{Eugene Demler} 
\affiliation{Department of Physics, Harvard  University, Cambridge, Massachusetts 02138, USA}

\author{Richard Schmidt} \thanks{
Corresponding author.
\\Email:  richard.schmidt@mpq.mpg.de.} 
\affiliation{Max-Planck-Institute of Quantum Optics, Hans-Kopfermann-Strasse. 1, 85748 Garching, Germany}
\affiliation{Munich Center for Quantum Science and Technology (MCQST), Schellingstr. 4, 80799 M\"unchen, Germany}

\begin{abstract}
We consider the quench of an atomic impurity via a single Rydberg excitation in a degenerate Fermi gas. The Rydberg interaction with the background gas particles induces an ultralong-range potential that binds particles to form dimers, trimers, tetramers, etc. Such oligomeric molecules were recently observed in atomic Bose-Einstein condensates. In this work, we demonstrate with a functional determinant approach that quantum statistics and fluctuations have observable spectral consequences.  We show that the occupation of molecular states is predicated on the Fermi statistics, which suppresses molecular formation in an emergent molecular shell structure. At large gas densities this leads to spectral narrowing, which can serve as a probe of the quantum gas thermodynamic properties.
\end{abstract}

\maketitle

\section{Introduction}

The study of bound complexes composed of a large number of particles lies at the heart of physics, chemistry  and biology. Examples include DNA formed from nucleotides, complex molecules composed of atoms, and nuclei comprised of neutrons and protons. Our understanding of these complex systems emerges from  idealized models, which are yet sufficiently complex to contain the relevant physics. A prime example are shell models, which underlie our description of the structure of atoms \cite{Szabo}, nuclei \cite{NuclearBook} and quantum dots \cite{woggon1997}.  A key ingredient in shell models is the quantum statistics of the constituent particles, which is fermionic for  electrons, protons, and neutrons leading to Pauli exclusion and the concept of filled shells.

In nature, large bound complexes are often embedded in environments and reside in a state far from equilibrium. This presents an outstanding challenge for experiment and theory as now an understanding of the interplay of dynamics and quantum statistics in bound structures becomes essential to explain the evolution through the hierarchy of more complex structures as system size increases.

\begin{figure}[t]
\centering
\includegraphics[width=\columnwidth]{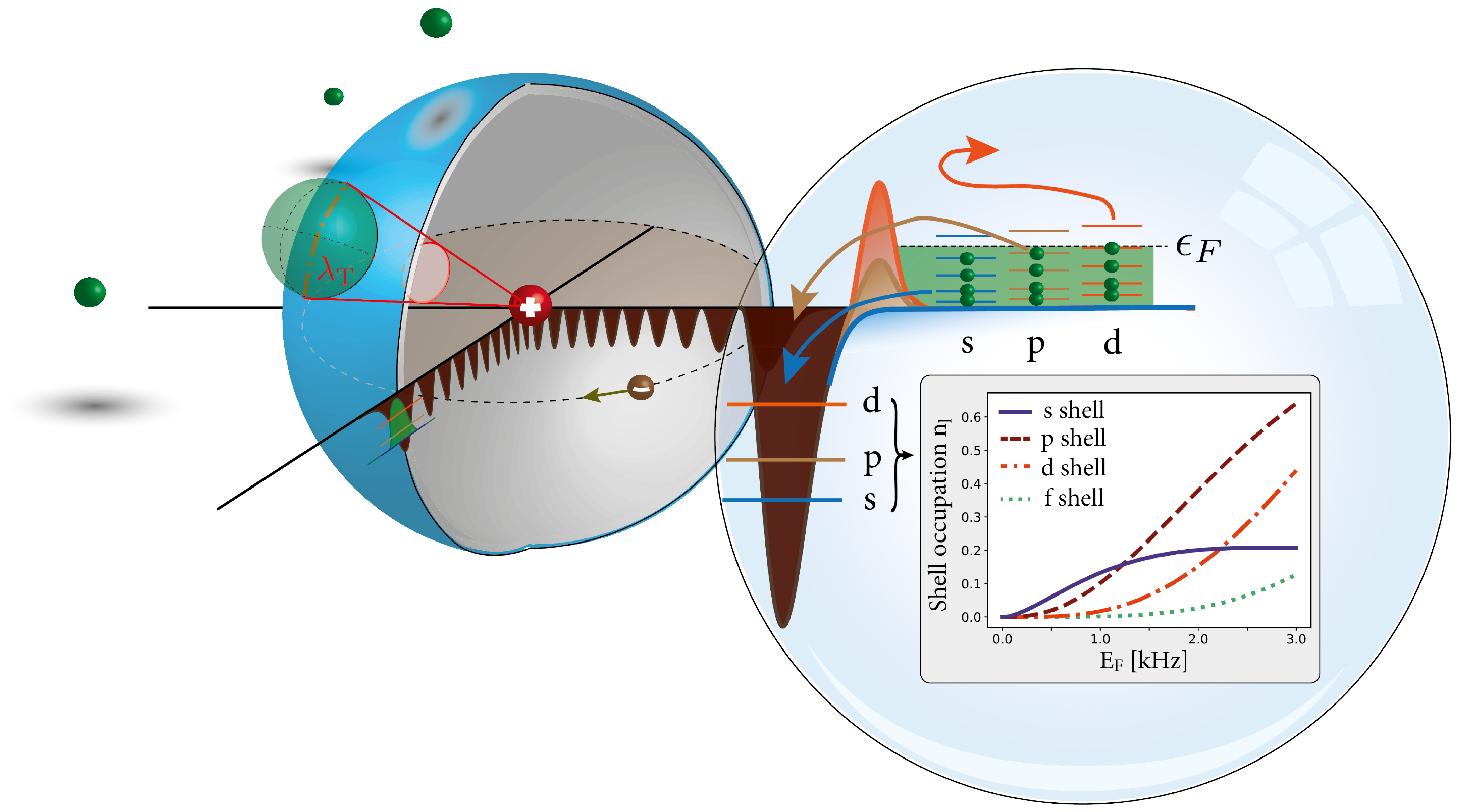}
\caption{\textbf{Occupation of a shell structure induced by a Rydberg excitation in a Fermi sea.} The Rydberg electron (brown sphere) of the impurity atom (red ion core) induces a molecular potential (dark brown) for the host atoms whose range can be tuned by  the principal quantum number $n$. As illustrated in the zoom-in  to the right (blue bubble), the  potential supports bound molecular states in various angular momentum channels that are localized in the outermost potential well (not to scale).  These bound states can be populated by the atoms initially occupying the single-particle states of the unperturbed Fermi sea up to the Fermi energy $\epsilon_F$ (green shading). Since angular momentum is conserved in the scattering with the spherically symmetric Rydberg excitation, fermions must overcome the rotational barrier to occupy the $s$, $p$ and $d$ shells of the Rydberg atom leading to a shell structure whose occupation (shown in the inset) is determined by a Pauli-enforced Rydberg rotational blockade. 
}
\label{fig1}
\end{figure}

Here we show that a Rydberg impurity interacting with a background atomic gas leads to the formation of a molecular shell structure in the quantum regime, where the thermal de-Broglie wave length $\lambda_T$ becomes comparable to the range of the impurity-background gas potential, see Fig.~\ref{fig1}. In contrast to other shell models, particles can  be bosonic or fermionic so that the role of quantum statistics can be explored. The shell structure we describe here for fermions is akin to the nuclear shell model and arises because at low gas density few-body bound molecular states must obey the Pauli exclusion, leading to a sequential filling of molecular shells.  At larger densities, Rydberg excitations in a Fermi gas are an example for systems where quantum mechanics and statistics can appear in surprising places:  The atoms bound in the molecular shells are localized on a length scale that is much smaller than the inverse Fermi momentum. One might thus expect that quantum statistics should be irrelevant. We find that this is not the case. With a functional determinant approach \cite{levitovFDA1,levitovFDA2}, we monitor the time evolution of the Fermi gas, subject to a sudden Rydberg excitation, in a superposition state of antisymmetrized many-body wave functions.  We show that an intricate interplay of wave function overlaps and quantum statistics leads to physics similar to the Anderson orthogonality catastrophe \cite{AndersonOC,nozieres} with an observable Fermi suppression of spectral density  containing the direct signature of many-body dressing.

The salient features of our results are:
 
\emph{\textbf{1.\,Mesoscopic Pauli exclusion: rotational blockade and inhibition of molecule formation.---}} Atoms bound in the Rydberg potential occupy states that can be labeled by vibrational and rotational quantum numbers $k$, $l$, and $m$. While an arbitrary number of bosons can occupy each of these levels, for fermions Pauli statistics manifests in the filling of a shell structure. Upon filling the $s$ ($l=0$) shell, the next three atoms  are placed in the $p$ ($l=1$) shell, and so on. We find that the occupation of the molecular shells from atoms in a low-density Fermi gas is rotationally blocked due to Pauli exclusion. This leads to a distinct response of Fermi and Bose gases, and a suppression of the formation of fermionic, ultralong-range  Rydberg molecules.

 \emph{\textbf{2.\,Fermi\,compression: many-body spectral narrowing and probe of macroscopic quantities.---}} At large gas densities, we find that the many-body spectral density is measurably narrower for fermions than for bosons. In Bose gases \cite{Rydbergpolaron1,Rydbergpolaron2,Rydbergpolaron3}, it was found that the large extent of the impurity potential and dressing with bound states leads to the emergence of a superpolaronic response which can be understood from the independence of bosonic modes. For fermions such a description breaks down. Instead, we find that the competition between Fermi pressure and bound-state formation leads to an observable compression of the absorption spectrum when compared to the response in a dense bosonic environment. Being tied to the local density fluctuations in the environment, this  spectral compression can serve as a novel local {\it in-situ} probe of pressure and compressibility in quantum gases. Since the Pauli pressure is the thermodynamic manifestation of the Fermi correlation hole, Rydberg impurities thus present a new and controllable tool to study the origin of thermodynamic properties on the microscopic scale.

The paper is organized as follows. In Section~\ref{RydMod}, we discuss the physical setup of a Rydberg excitation in an ultracold Fermi gas and introduce the relevant microscopic model. Then, in Section~\ref{Method}, we introduce a functional determinant approach that captures the quench dynamics of the Fermi sea in response to the sudden introduction of a gigantic impurity excitation. Section~\ref{RydShellStructure} demonstrates how a shell structure emerges at various densities.  There, we show that a Pauli-enforced Rydberg rotational blockade leads to suppressed molecular formation at low densities and  a compressed superpolaronic excitation at higher densities. We relate the predicted phenomena to the orthogonality catastrophe (OC)  in Section~\ref{OC}. We conclude the article in Section~\ref{Conc} by providing an outlook and outlining experimental protocols to measure the predicted phenomena.

\section{A Rydberg impurity in a Fermi sea} \label{RydMod}
We consider a single Rydberg atom suddenly immersed in a degenerate Fermi gas of spin-polarized ultracold atoms at temperature $T$ and homogeneous particle density $\rho$. A limiting case of our theory is the $T=0$ ensemble, in which the initial state of the many-body environment is given by $\ket{\Psi_{\rm FS}} = \prod_{|{\bf k}| \leq {k}_{\rm F}}  \hat{c}^\dagger_{\bf k} \ket{0}$, where fermions of mass $m$, described by creation operators $\hat{c}^\dagger_{\bf k}$, fill  single-particle orbitals of momentum $\veck$ up to the Fermi momentum $k_{\rm F}$. \\

\noindent\textbf{\textit{Rydberg impurity-bath interaction.---}} Upon excitation to a state $\psi_e(\bf{r})$ of principal number $n$, the Rydberg electron interacts with the ground-state atoms in its environment. The frequent scattering of the low-energy Rydberg electron from the gas perturber atoms was first described by Fermi \cite{FermiPseudoPotential,RydMoleculeReview}, and leads to a Born-Oppenheimer interaction potential between the Rydberg and ground state atoms, 
\begin{equation} \label{eqn:RydPot}
V_{{\rm Ryd}} ({\bf r}) = \frac{2 \pi \hbar a_e}{m_e} |\psi_e({\bf r})|^2.
\end{equation}
Here $a_e$ is the electron--ground-state-atom scattering length, $m_e$ the electron mass, and  $\mathbf r$  the distance separating a ground-state atom from the ionic core of the Rydberg impurity. 

\begin{figure}[t]
\centering
\includegraphics[width=\columnwidth]{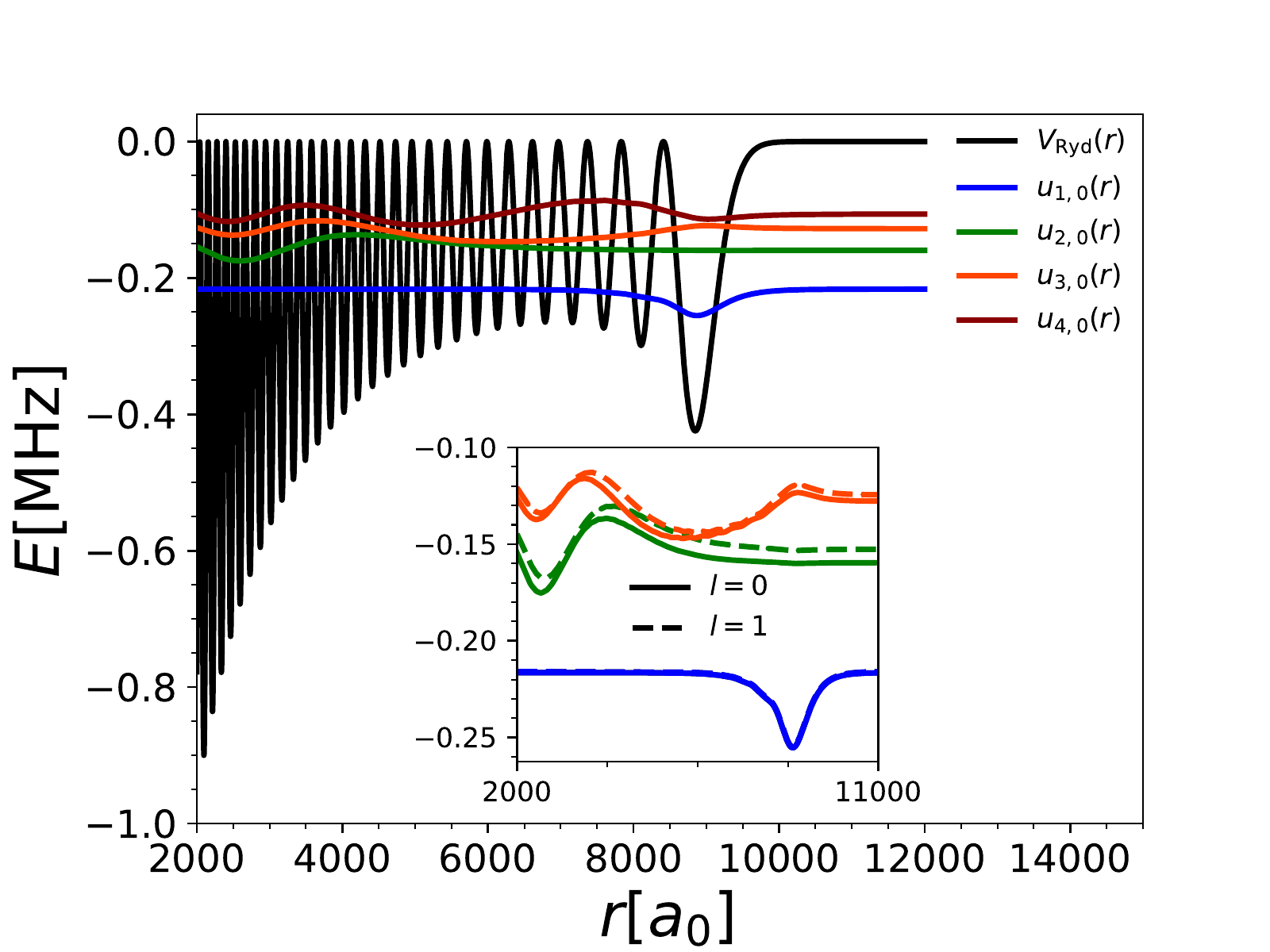}
\caption{\textbf{Rydberg molecular potential.} The Rydberg potential $V_{\rm Ryd}(r)$ for a $^{87}$Rb$(71s)$ excitation as obtained from a quantum defect calculation in an $s$-wave scattering approximation. The potential is shown as a function of the distance $r$ between the Rydberg ion and an atom in the Fermi sea. The interacting single-particle bound radial wave functions $u_{k_{\alpha},l_{\alpha}}(r)$ are shown as colored lines. The inset shows the lowest three vibrational states for angular momenta $l=0$ (solid) and $l=1$ (dashed). The offset of the wave functions corresponds to their respective energies. This exemplary potential is used for all numerical results shown in this work.
}
\label{fig2}
\end{figure}

The oscillatory nature of the potential, shown as the black line in Fig.~\ref{fig2}, reflects the nodal structure of the Rydberg electron wave function $\psi_e(\bf{r})$ \cite{Rydbergpotential}. For $a_e<0$, this potential supports bound vibrational states \cite{RydbergmoleculesTheo}, see Fig.~\ref{fig2}. Since the principal number $n$ can be chosen in experiments, the scaling of the range and oscillations of $|\psi_e({\bf r})|^2$ with $n$  offers a complimentary tool for the control of interactions beyond the widely employed magnetically or optically tuned Fano-Feshbach resonances \cite{Feshbachgases}.\\

\noindent\textbf{\textit{Rydberg impurity model.---}} Rydberg impurities in Fermi gases can be realized in atomic mixtures where a small fraction of one species is immersed in a Fermi gas of another species. A laser resonant with an atomic transition of the minority species excites a single atom from its electronic ground state denoted as $\ket{\downarrow}$  to a highly excited Rydberg state $\ket{\uparrow} = \ket{n}$ (here we suppress all other quantum numbers), creating a single impurity in the Fermi gas.

To keep the analysis transparent, we focus  on heavy Rydberg impurities, for which the ion recoil can be  neglected.  Furthermore, at low temperatures direct fermion-fermion interactions can be completely ignored. To this end, the Rydberg impurity interacts with the degenerate Fermi gas according to the Hamiltonian:
\begin{equation} \label{eqn::Ham}
\hat{H} = \sum_{\bf{k}} \epsilon_{\bf{k}} \hat{c}^\dagger_{\bf{k}} \hat{c}_{\bf{k}} + {\frac{1}{\mathcal{V}}}\sum_{\bf{k},\bf{q}}V_{\rm Ryd}({\bf{q}}) \hat{c}^\dagger_{\bf{k}+\bf{q}} \hat{c}_{\bf{k}} \hat{d}^\dagger \hat{d},
\end{equation}
where $\mathcal{V}$ is the system volume, and $\hat{d}$ is the annihilation operator of the heavy Rydberg impurity atom in the $\ket{\uparrow}$ state initially localized at ${\bf R} = 0$. It interacts with bath fermions with dispersion relation $\epsilon_{\bf{k}} = {\bf k}^2/2m$ via the potential $V_{\rm Ryd}({\bf q})$, the Fourier transform of Eq.~(\ref{eqn:RydPot}), which in real space takes the form $\int_{\bf r} d{\bf r}\hat{c}^\dagger({\bf r}) \hat{c}({\bf r}) V_{\rm {Ryd}}({\bf r}) \hat{d}^\dagger \hat{d}$. In this work we compare the physics of Rydberg excitations in bosonic and fermionic environments. To allow for such a direct comparison all numerical results are shown for  bath atoms (fermionic or bosonic) of atomic mass 87, and a representative potential $V_\text{Ryd}(\vecr)$ as calculated for a $^{87}$Rb$(71s)$ Rydberg excitation; see Fig.~\ref{fig2}.

\section{Fermionic many-body dynamics induced by Rydberg excitations} \label{Method}

Previous theoretical analyses of the interaction of Rydberg with ground-state atoms and the resulting molecule formation have mostly focussed on calculations of binding energies, wave functions, and density-independent spectral lines of dimers or timers \cite{RydFewBodyTheo1,RydFewBodyTheo2,RydFewBodyTheo3,BoothRydMol,Rydbergimpurexp3,RydMoleculeReview,RybMolAnis}. Recent work \cite{Rydbergpolaron1,Rydbergpolaron3} introduced an approach to study the many-body quantum dynamics of bosonic systems. Here, we extend this method to fermions by combining an atomic physics few-body approach to Rydberg molecule formation with many-body techniques of mesoscopic physics, which allows us to accurately capture the multiscale nature and non-perturbative character of the Rydberg impurity problem.\\

\noindent\textbf{\textit{Absorption line shapes from quench dynamics.---}} The frequency-resolved absorption spectrum $A(\omega)$ is obtained from the Fourier transform 
\begin{equation}\label{eqn::AS}
A(\omega) = 2 {\rm Re} \int_0^\infty dt e^{i\omega t} S(t)
\end{equation}
of the time-dependent overlap function \cite{Mahan,RamseyOC2,FermiPolaronRevRich} (Appendix \ref{AbsS})
\begin{equation}\label{eqn::LE}
S(t) = {\rm Tr} \left[  e^{i\hat{H}_0t}  e^{-i\hat{H}t} \hat{\mathcal{\varrho}} \right],
\end{equation}
where $\hat{\mathcal{\varrho}} = e^{-\beta (\hat{H}_0 - \mu \hat{N})}/Z$ is the density matrix  of free fermions in absence of the impurity at inverse temperature $\beta = 1/k_BT$ and chemical potential $\mu$, with $k_B$ the Boltzmann constant and $Z$ the partition function of the Fermi gas. The Hamiltonian in the absence of the impurity is given by $\hat{H}_0$, and $\hat{H}$ is the Hamiltonian in the presence of the impurity [Eq.~\eqref{eqn::Ham}]. The expression Eq.~\eqref{eqn::LE}, also known as the Loschmidt echo, describes the dephasing dynamics of the fermionic environment following a quench of the impurity-bath potential due to the sudden  introduction of the Rydberg excitation in the atomic gas. It can be directly measured  using Ramsey spectroscopy \cite{RamseyOC1,RamseyOC2,fdg3,fdg4,Parish2016,Mistakidis2019}. \\

\noindent\textbf{\textit{Functional determinants.---}} To compute the  quantum dynamics $S(t)$, we employ a functional determinant approach (FDA) \cite{levitovFDA1,levitovFDA2}, which provides exact numerical results for systems described by bilinear Hamiltonians. The strength of the FDA lies in the ability to reduce expectation values of many-body operators to determinants in the single-particle Hilbert space (for details see Appendix \ref{FDA_Appendix}), taking into account infinitely many bath excitations. In this way, the FDA allows one to efficiently compute the many-body dynamics induced by the Rydberg impurity in the Fermi sea, keeping track of the full antisymmetrization of the many-body wave functions, along with the Boltzmann factors needed for thermal averaging at finite temperatures.

Within the FDA, the time-dependent overlap, Eq.~\eqref{eqn::LE}, evaluates to (Appendix \ref{SPS})
\begin{equation}\label{FDA}
S(t) = {\rm{det}}[1 - \hat{n}_{{\rm FD}} + \hat{n}_{{\rm FD}} e^{i\hat{h}_0 t}e^{-i\hat{h}t}],
\end{equation}
where the determinant is calculated in the single-particle Hilbert space, with $\hat{h}$ and $\hat{h}_0$ being the one-body counterparts of $\hat{H}$ and $\hat{H}_0$, respectively.  Furthermore, $\hat{n}_{{\rm FD}}= \frac{1}{e^{\left( \hat{h}_0-\mu \right)/k_B T} + 1}$ gives the Fermi-Dirac distribution in the `non-interacting' single-particle orbitals determined by $\hat h_0$.

Since the Rydberg potential is spherically symmetric,  different angular-momentum $l$ channels are decoupled. As a result, the Loschmidt echo factorizes into a product of $l$-dependent terms (Appendix \ref{SPS})
\begin{eqnarray} \label{SFSS}
S(t) = \prod_{l} S^{l}(t),
\end{eqnarray}
where $S^{l}(t) = [s_{l} (t)]^{2l+1}$ encodes the quantum dynamics within the given $l$ manifold with $2l+1$ degenerate $m$ states, and 
\begin{align} \label{LE_lm}
s_{l}(t)  = &\quad {\rm det}\Big[ \delta_{s,s'} \Big(1- n_{\rm FD}(\epsilon_{s})\Big) \nonumber \\
 & + n_{\rm FD}(\epsilon_{s'}) \sum_{k_{\alpha}} e^{i(\epsilon_{s'}-\omega_{\alpha})t}\bra{k_{s'}}\ket{k_{\alpha}} \bra{k_{\alpha}}\ket{k_s}\Big].
\end{align}

Here, $\alpha$ and $s$ are collective indices that include the nodal quantum number $k$, angular momentum $l$ and projection $m$ of  the interacting and non-interacting single-particle eigenstates, respectively. To evaluate this expression, we calculate the radial wave functions in presence of the impurity, $u_{k_{\alpha}}(r) = \bra{r}\ket{k_{\alpha}}$, and in its absence, $u_{k_{s}}(r) = \bra{r}\ket{k_{s}}$, of eigenenergies $\epsilon_{\alpha}$ and $\epsilon_s$, respectively.  The single-particle orbitals are obtained numerically from the bound and continuum eigensolutions of the Schr\"{o}dinger equation for a localized Rydberg impurity, for details  see Appendix \ref{SPS}.

\section{Pauli-enforced rotational blockade and Fermi compression}\label{RydShellStructure}

In the following, we  expand on the discussion of the two main features of this work: a) although the angular momentum shells have a typical spacing of $\sim1$KHz (determined by the Rydberg molecule rotational constant) and are thus not spectroscopically resolved, it is the Franck-Condon overlaps of bound molecular wave functions with the free single-particle wave functions, that ultimately determine the spectral intensity.  Those overlaps are exponentially suppressed, see~Fig.~\ref{fig6}, and lead to the inhibition of fermionic molecule formation. We term this effect rotational blockade; b) the quantum statistics of Fermi occupation leads at higher gas densities to a spectral narrowing of superpolaronic features which we interpret as a Fermi compression.

To appreciate the distinct properties of Rydberg impurities in a fermionic many-body environment, it is instructive to first contrast their physics to the recently experimentally realized scenario \cite{FermipolaronExpChevySol,koschorreck_attractive_2012,RepulsivePolaronFermiExp,Zhang2012,RepulsivePolaronFermiExp,fdg1,fdg3,fdg4} of  impurities interacting with  bath atoms via attractive contact interactions. There, if the interaction potential is sufficiently attractive, dimers can form. Those dimers exist, however, only in a state of zero-angular momentum. This is in stark contrast to Rydberg impurities in a Fermi gas where, due to the large extent of the impurity-bath interaction, Rydberg molecules form in states of finite angular momentum. This difference precludes cold atoms interacting solely by short-range interactions as a platform to study the competition of Pauli exclusion and occupation of bound states of higher angular momentum,  essential to realize the physics of shell structures, rotational blockade and Fermi compression.

\begin{figure}
\includegraphics[width=\columnwidth]{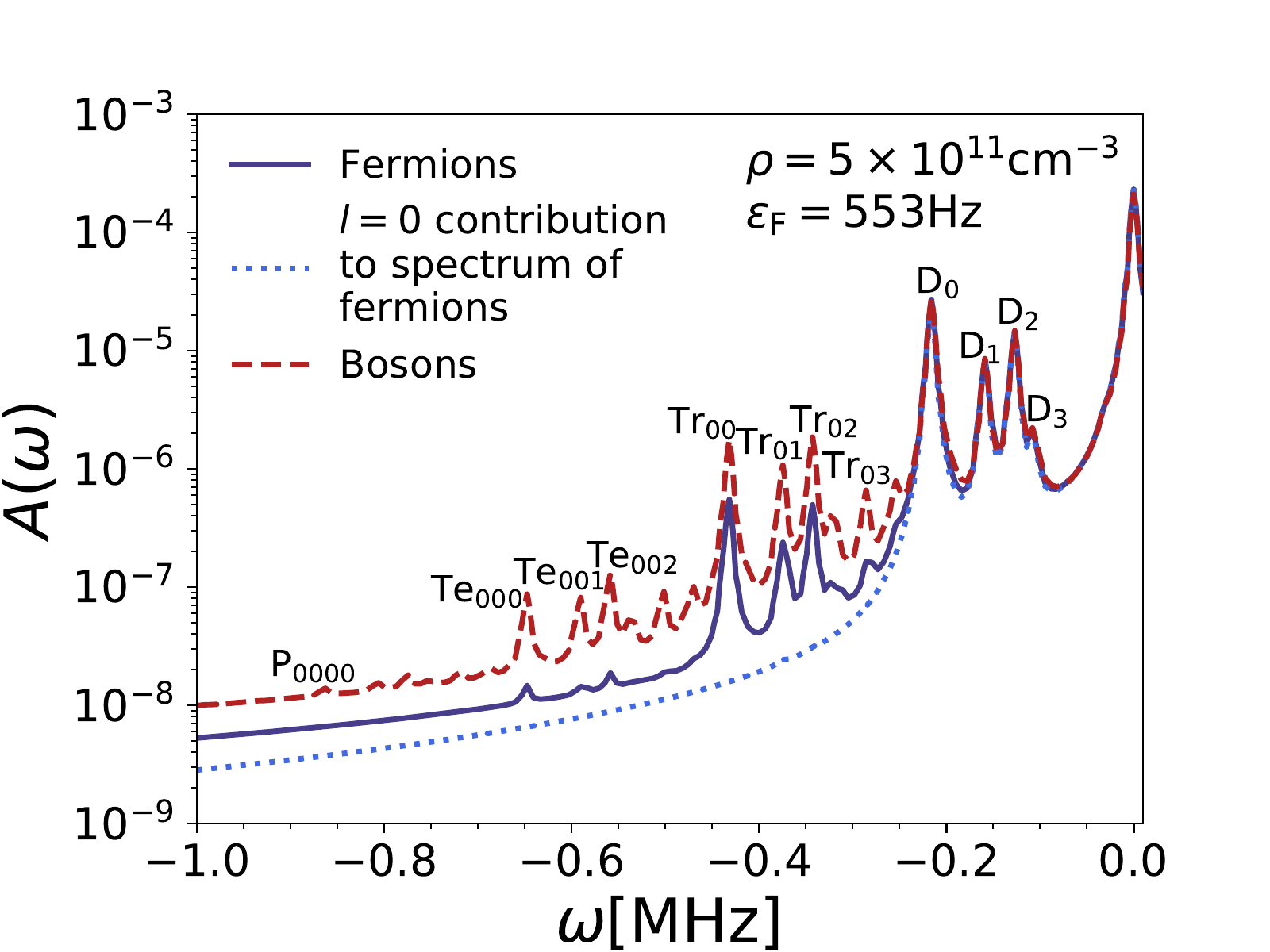}
\caption{\textbf{Impact of quantum statistics at low density.} Absorption spectrum $A(\omega)$ of the Rydberg impurity in the Fermi sea (solid line) and its $l=0$ component (dotted line), compared to the spectrum obtained for a Rydberg excitation in a BEC (dashed line) at  low density, $\rho = 5\times10^{11}{\rm cm}^{-3}$.}
\label{fig5}
\end{figure}

\subsection{Absorption spectrum at low densities}
In Fig.~\ref{fig5}, we compare the spectrum of a Rydberg impurity in a Fermi sea to that in a BEC, both at  low density $\rho =5\times10^{11}$ cm$^{-3}$. The fermionic response (solid blue) is calculated according to Eq.~\eqref{SFSS}, while the spectrum for the BEC (red dashed) is obtained using the methods described in Ref. \cite{Rydbergpolaron1}, see also Appendix \ref{StBEC}.

\begin{figure*}
    \includegraphics[width=\columnwidth]{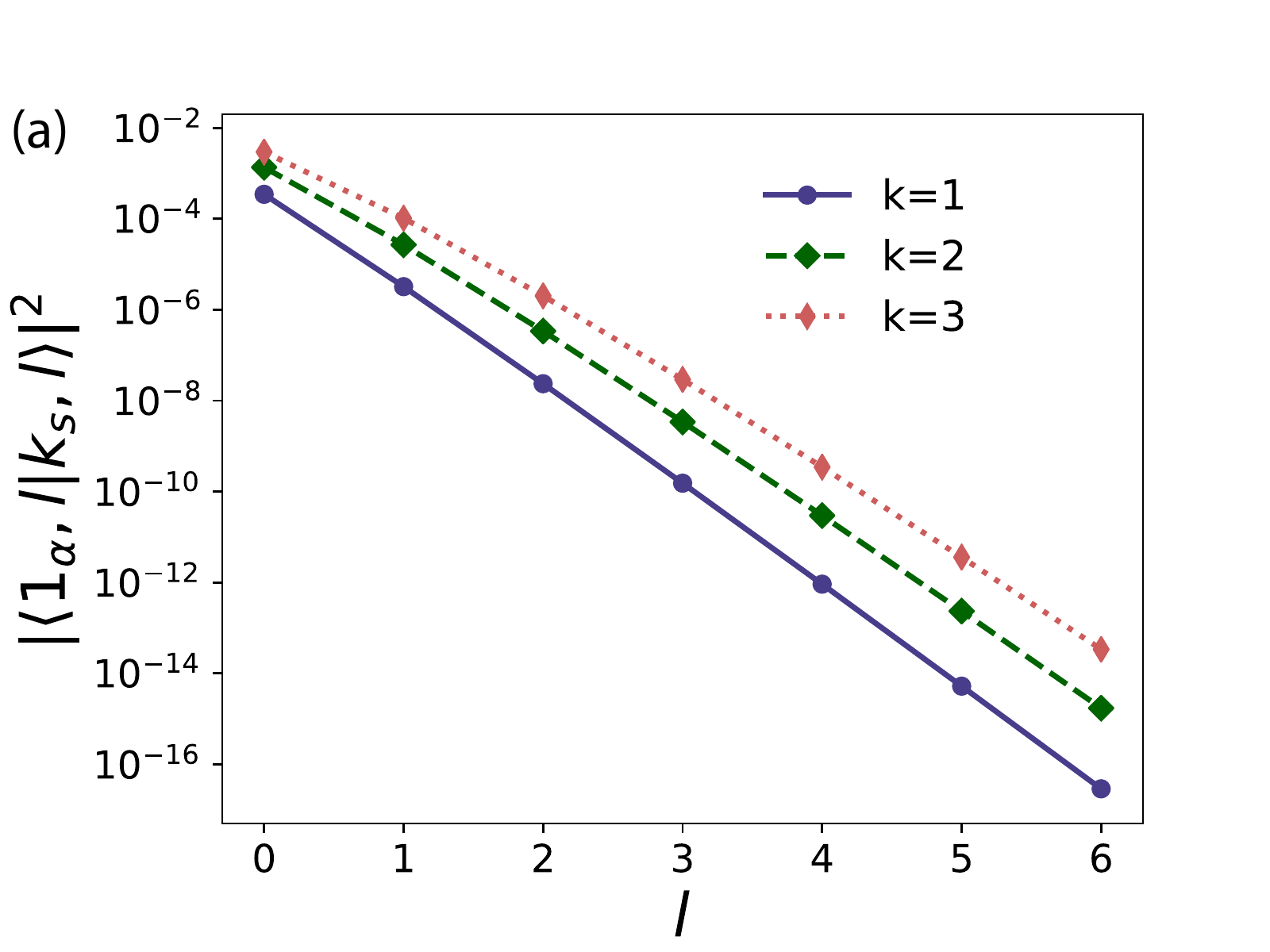} \includegraphics[width=\columnwidth]{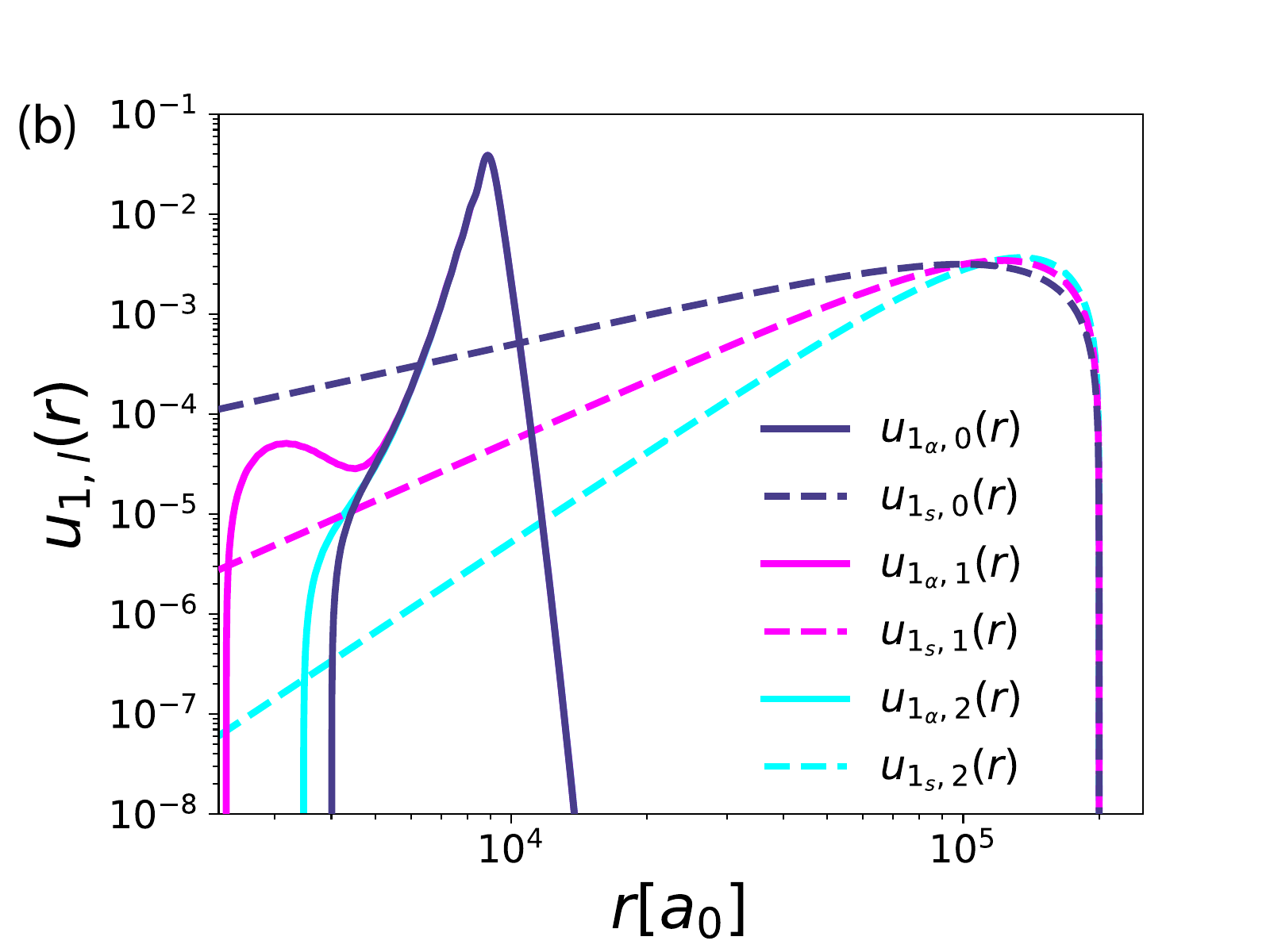}
\caption{\textbf{Rydberg rotational barrier.} 
(a) Franck-Condon factors $\left|\bra{1_\alpha,l}\ket{k_s,l}\right|^2$ between the lowest molecular dimer states $\ket{1_\alpha,l}$ and the non-interacting states $\ket{k_s,l}$ of nodal number $k=1$, $2$ and $3$ as a function of the rotational angular momentum $l$. (b) Comparison of the spatial structure of the interacting and non-interacting single-particle states that give rise to the  super-exponential decay with $l$  of the Franck-Condon factors shown in panel (a). In contrast to the non-interacting states (dashed lines), the molecular dimer states (solid lines) in the $s$, $p$ and $d$ orbitals of the shell structure are deeply localized in the outermost potential well and are thus hardly affected by the rotational barrier. }
\label{fig6}
\end{figure*}

Even at such a low density, we observe a suppression of spectral weight at large detuning $\omega$ when comparing the fermionic response to that of a BEC. While dimers, D$_0$, D$_1$, D$_2$, $\cdot\cdot\cdot$, form both in the Fermi  and Bose gas with the same line strength, higher-order molecular complexes, such as trimers, Tr$_{00}$, Tr$_{01}$, Tr$_{02}$, $\cdot\cdot\cdot $, tetramers,  Te$_{000}$, Te$_{001}$, $\cdot\cdot\cdot$, and pentamers, P$_{0000}$, $\cdot\cdot\cdot$, are suppressed (the indices denote the vibrational quantum numbers of the  molecular states involved).

As in the nuclear shell model, the bound-state configurations must obey the Pauli exclusion principle. Thereby trimers, tetramers, {\em etc}$\ldots$ are composed of particles in different angular $l$ and magnetic $m$ orbitals. As such, Rydberg impurities provide a cold-atom analogue of the nuclear shell model \cite{NuclearBook},  with orbitals that can be adjusted by choosing the principal number of the Rydberg  excitation. In contrast to nuclear physics \cite{NuclearBook},  in the cold-atom setting the confining potential is not self-generated by the particles that become bound, but by the interaction between the Rydberg electron and background atoms. Therefore, it becomes possible to switch on or off the shell-model-defining potential, and on timescales much faster than the inverse binding energies. This also implies that for Rydberg excitations the formation of the shell structure is inherently \textit{dynamical}. This allows studying the full spectrum of the model and the non-equilibrium occupation of its excited shells.

The shell structure concept helps to unravel the mechanisms at play in Fig.~\ref{fig5}. While the introduction of the Rydberg  impurity modifies the single-particle spectrum of the background gas by introducing bound states and affecting the continuum scattering states by small phase and energy shifts, it conserves the total angular momentum. Using this fact we decompose the response shown in Fig.~\ref{fig5} in the various $l$ channels, and show the contribution to the fermionic spectrum from atoms in states of angular momentum $l = 0$. Evidently, this contribution (blue dotted line) accounts for most of the overall absorption response (solid line). In simple terms, this effect can be understood from the fact that atoms occupying $l=0$ states in the initial Fermi sea are the only ones with substantial spatial overlap with the volume of the Rydberg excitation. While the $l=0$ contribution accounts for nearly all of the fermionic dimer response,  trimer and higher-order lines are  missing. On the one hand this suppression arises since $l=0$ states can only be occupied once. On the other hand, while this argument prohibits the formation of a Tr$_{00}$ trimer, it does not exclude the trimers such as Tr$_{01}$, that are constituted of atoms occupying different vibrational states, akin to the occupation of higher vibrational orbitals in the nuclear shell model.  Compared to the bosonic environment, even these trimer and tetramer states are, however, suppressed.  The reason for this finding is that by the restriction to the $l=0$ subspace one does not account for the total number of fermions in the system. Thus, the isolated $l=0$ contribution corresponds to an effective density that is lower compared to the $T=0$ BEC, where {\em all} bosons reside in the zero angular momentum subspace.

The remaining contribution to the full spectrum in Fig.~\ref{fig5} (solid line) originates thus from fermions that initially resided in finite angular momentum states of the non-interacting Fermi sea.  Due to the conserved angular momentum, these fermions can occupy states in the molecular shell structure that have the same final angular momentum. Considering the extremely small rotational constant of the Rydberg molecules, a large number of such rotational shells is, in principle, energetically available to the bath atoms. Thus one might expect the spectral response of fermions and bosons to be similar. Consequently, differences due to occupation of higher $l$ states would only be observable if the rovibrational energies become comparable to the experimental resolution (which may be in reach for lighter atomic species such as $^6$Li). Considering this argument it is thus at first surprising that in Fig.~\ref{fig5} quantum statistics does apparently plays a role in determining the absorption response of fermions compared to that of bosons.

The puzzle is resolved when considering the collisional many-body physics involved: the initial state of the Fermi sea describes atoms that fill non-interacting single-particle orbitals in various angular momentum modes up to the Fermi level (for an illustration see Fig.~\ref{fig1}).  The higher the angular momentum of those states, the smaller the spatial overlap of their single-particle wave function with the volume of the Rydberg excitation becomes. This leads to the observation that while the angular-momentum orbitals are indeed energetically quasi-degenerate, the Frank-Condon factors (FCFs) between non-interacting and interacting single-particle states exhibit systematic variations with angular momentum that help to explain the absorption response.

In Fig.~\ref{fig6}(a), we illustrate the $l$-dependence of the single-particle FCFs (or overlaps) $\left|\bra{1_{\alpha},l}\ket{{k_s,l}}\right|^2$ between the lowest non-interacting  states $\ket{{k_s,l}}$ ($k=1$, $k=2$ and $k=3$) and the bound dimer $\ket{{1_{\alpha},l}}$.  A super-exponential decay with  $l$ of the FCFs, which characterize the probability for occupying bound states from the initially non-interacting state of free fermions, is found (see Appendix~\ref{Analytic}). This decay  can be traced to the centrifugal angular-momentum barrier and the suppression of the non-interacting single-particle wave functions with higher $l$, within the volume of the  interacting bound state wave functions, see Fig.~\ref{fig6}(b). As can be seen there, the interacting wave functions are nearly the same for different $l$ as they experience an effective potential that is dominated by the Rydberg interaction. In contrast, the effective potential for the free fermions is solely determined by the centrifugal barrier, which results in non-interacting wave functions that are increasingly suppressed at small distances with higher $l$. We refer to the suppression of absorption response as {\em Pauli-enforced Rydberg rotational blockade} since it is the Pauli principle that forces additional particles that could be bound within the Rydberg orbital to occupy higher $l$ single-particle states, thus suppressing  the FCFs. The Rydberg rotational blockade of the Rydberg molecular shell structure is also evident when considering the average occupation number $n_l$ of the various $s$, $p$, $d$ shells in the Rydberg orbit. Their occupation, shown as inset in Fig.~\ref{fig1}, requires atoms in the initial state to overcome the rotational barrier and thus demands a sufficiently high Fermi energy $\epsilon_F$.

We note that  rotational blockade effects are familiar from the collisional physics of ultracold atoms close to their electronic ground state \cite{BlockadeAtomic1,BlockadeAtomic2,DeMarco2019}. There, however, the range of the potential is determined by the van der Waals length, $\sim100$a$_0$, and the rotational barrier is therefore at much higher energies. In contrast, Rydberg molecules are created at much larger separations of $\sim1000$a$_0$, allowing for the physics of rotational blockade to be studied in a previously inaccessible parameter regime.

\subsection{Spectral evolution with increasing density}

\begin{figure}[t]
\includegraphics[width=\columnwidth]{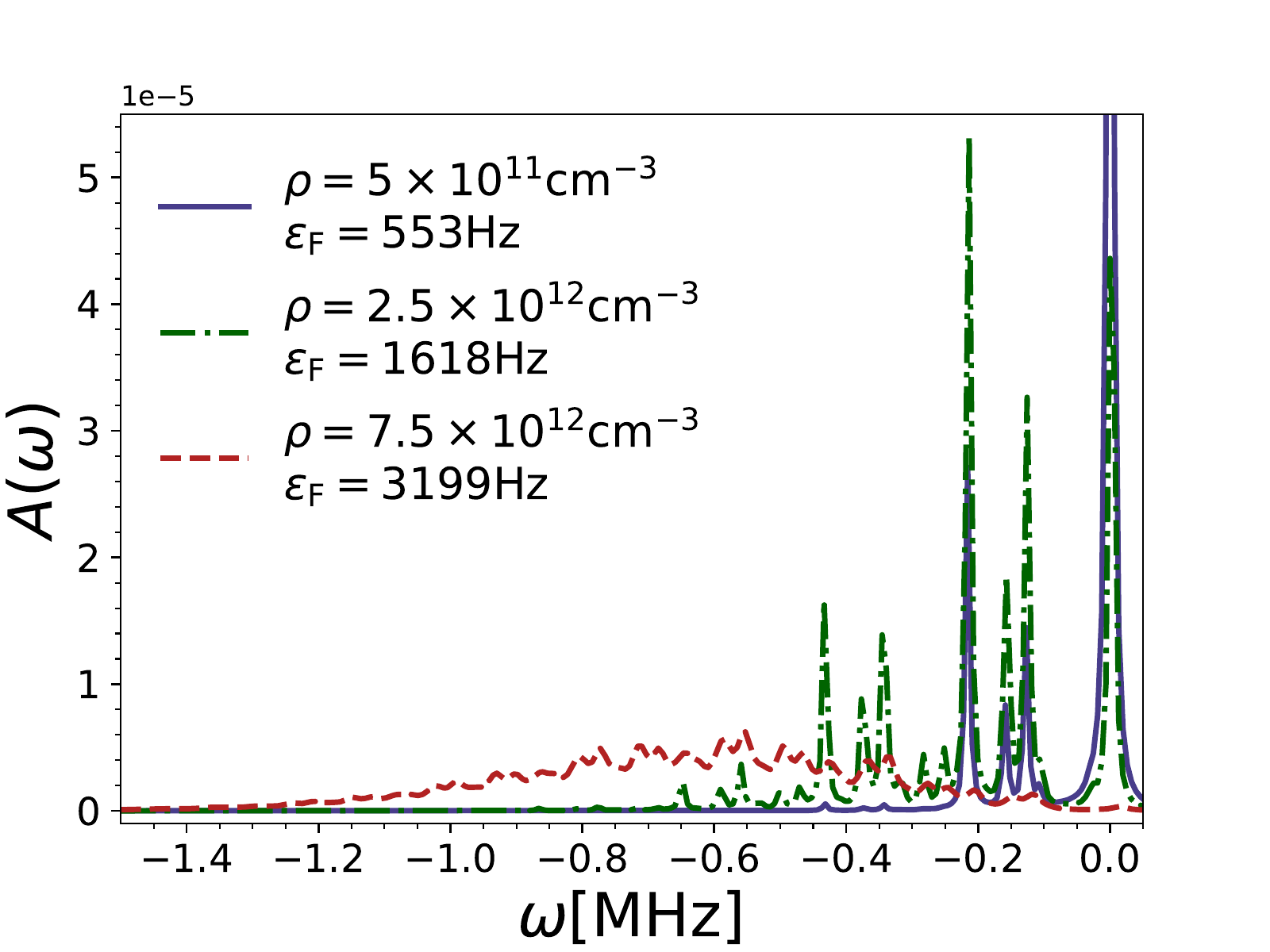}
\caption{
\textbf{Spectral evolution with density.} Absorption spectrum $A(\omega)$ of a Rydberg impurity in a Fermi sea of density $\rho =  5\times 10^{11} {\rm cm}^{-3}$, $2.5\times10^{12} {\rm cm}^{-3}$ and $7.5\times10^{12} {\rm cm}^{-3}$. $A(\omega)$ exhibits an evolution with increasing densities to a broad distribution as characteristic of the Fermi superpolaron. }
\label{fig3}
\end{figure}

We now turn to the behavior of the absorption spectra with increasing density.   In Fig.~\ref{fig3}, we show spectra for three densities corresponding to an increasing Fermi energy $\epsilon_{\rm F}$ that ranges from about $0.5$KHz to $3$KHz. As before, the low-density response (solid line) is dominated by the formation of few-body bound states. These result in a series of molecular lines that correspond to one or two background atoms bound inside the Rydberg orbit. As the density grows, a larger number of fermions occupy the bound states, see the dot-dashed line in Fig.~\ref{fig3}. For this density, corresponding to $\epsilon_{\rm F} \approx 1.6$KHz, the average inter-particle distance in the medium is $\sim 8700$a$_0$. From this, one estimates that on average about one atom is situated inside the Rydberg orbit which for the excitation considered in this work has the radius $R_0 \approx 8800$a$_0$.  In a classical statistical picture one would thus assume that, on average, a single atom binds to the impurity to form a dimer. In the setup considered here, the dimer has an energy $E_{\rm D} \approx 220 $KHz, and hence the mean of the spectrum is expected to occur at this frequency, which is consistent with the result from the full quantum calculation shown in Fig.~\ref{fig3}.

\begin{figure}[t]
    \centering
    \includegraphics[width=\columnwidth]{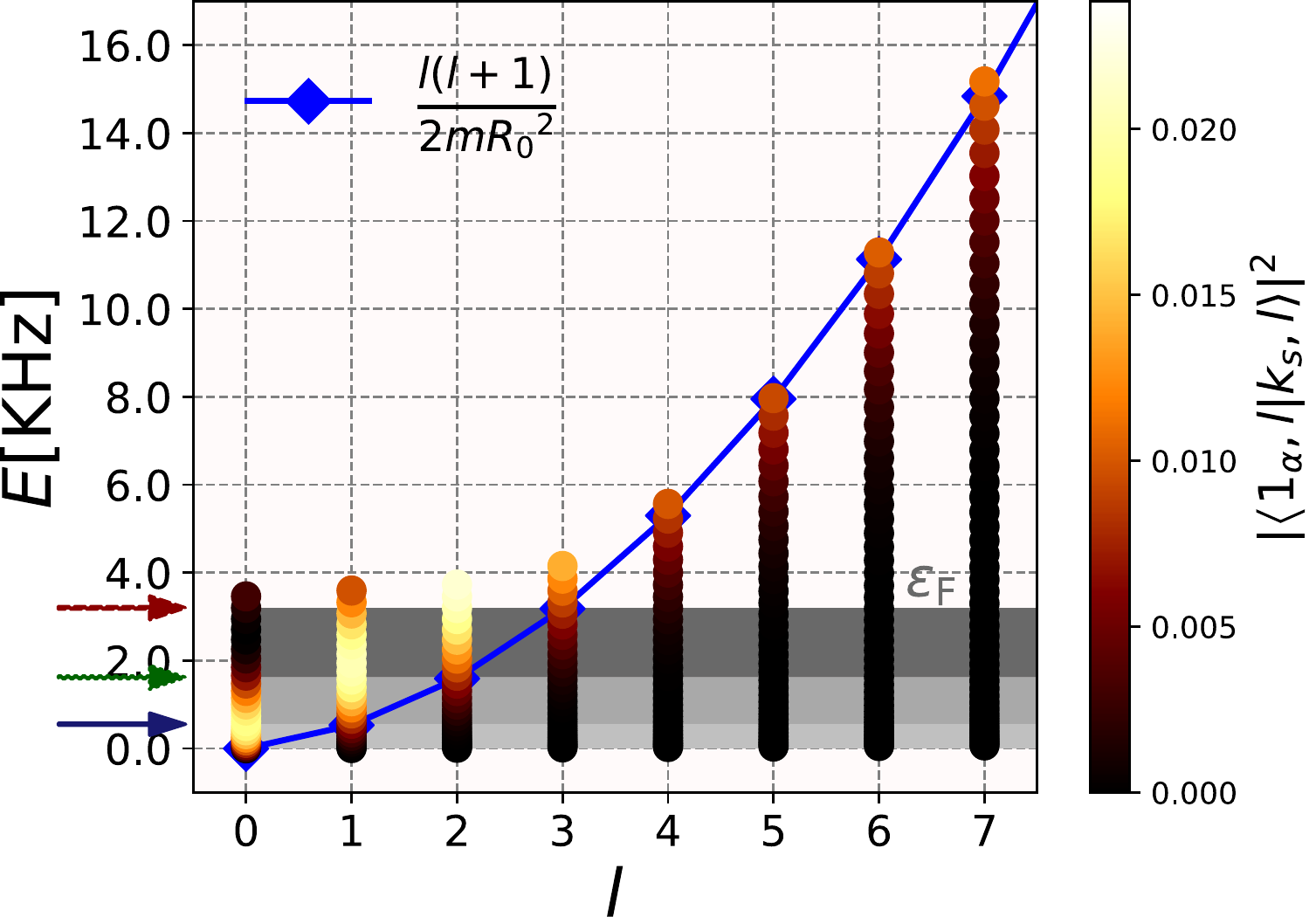} 
\caption{
\textbf{Effect of the rotational barrier on the initial Fermi sea.} The height of the centrifugal barrier $E_l(R_0)=l(l+1)/2 m R_0^2$ (blue line) at the radius $R_0$ of the outermost potential well, and the evolution with energy of the Franck-Condon overlaps $\left|\bra{1_\alpha,l}\ket{k_s,l}\right|^2$ between the lowest interacting nodal state $\ket{1_\alpha,l}$ and the non-interacting states $\ket{k_s,l}$ of nodal number $k$.  The arrows indicate the Fermi energies chosen for the calculations of the spectra  in Fig.~\ref{fig3}.
}
\label{fignewRep}
\end{figure}

The Fermi energy also determines which atoms from the initial state can, in principle, participate in the non-equilibrium Rydberg quench dynamics:  it is all of those atoms that have sufficient kinetic energy (including atoms with angular momentum $l=0$) to overcome the rotational barrier and acquire appreciable overlap with the bound molecular wave functions. In Fig.~\ref{fignewRep}, we show the energies of the non-interacting single-particle states (filled, colored circles) in the absence of the Rydberg impurity and the rotational barrier $\sim l (l+1)/2mR_0^2$ at the radius $R_0$ of the Rydberg electron orbit (solid blue line). In the color code, we  show their corresponding Franck-Condon factors (FCFs) with the lowest Rydberg molecular dimer state. The evolution of the FCFs with energy for the various $l$ states shows that the dimer states become rapidly accessible to the Fermi atoms once their energy crosses the rotational barrier. For instance, for a  Fermi energy $\epsilon_{\rm F}\approx1.6$KHz, see Fig.~\ref{fignewRep},  atoms in the $l=0$, $1$, $2$ states can overcome the barrier and participate in the dynamics. Combining this finding with the concept of the filling of shell structures now helps to explain the  corresponding absorption response shown as green dot-dash curve in Fig.~\ref{fig3}. Angular momentum is conserved in the bound state formation. Thus, although many  atoms are available in the $l=0,1,2$ initial states, at most only nine can actually become bound in the available molecular states since then the molecular shells of $l=0,1,2$ become filled as dictated by the Pauli exclusion principle. This is in stark contrast to the case of bosons, for which no bound on the rotational shell occupancy exists.

\begin{figure}[t]
    \centering
    \includegraphics[width=\columnwidth]{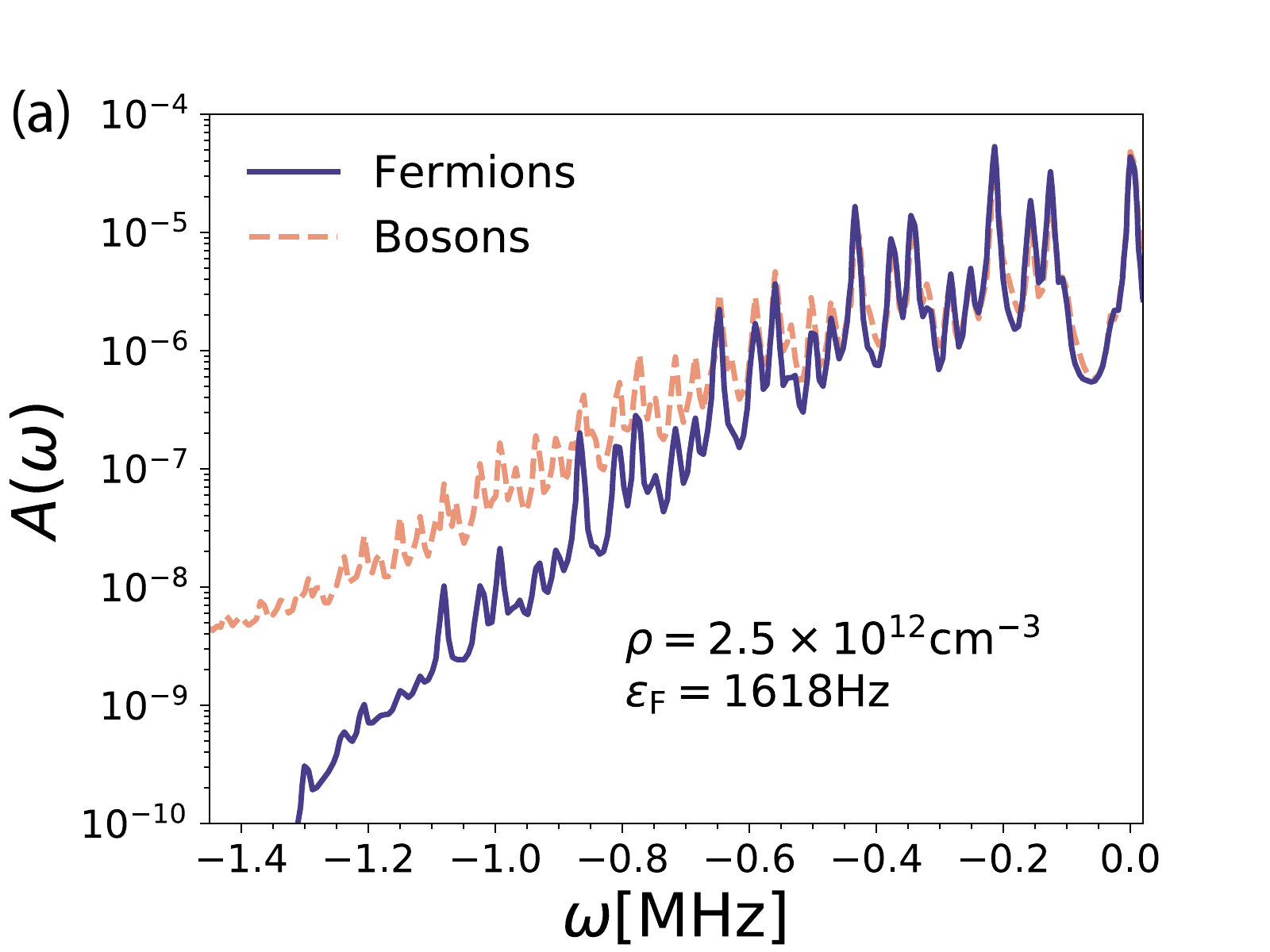} \newline\newline
    \centering
\includegraphics[width=\columnwidth]{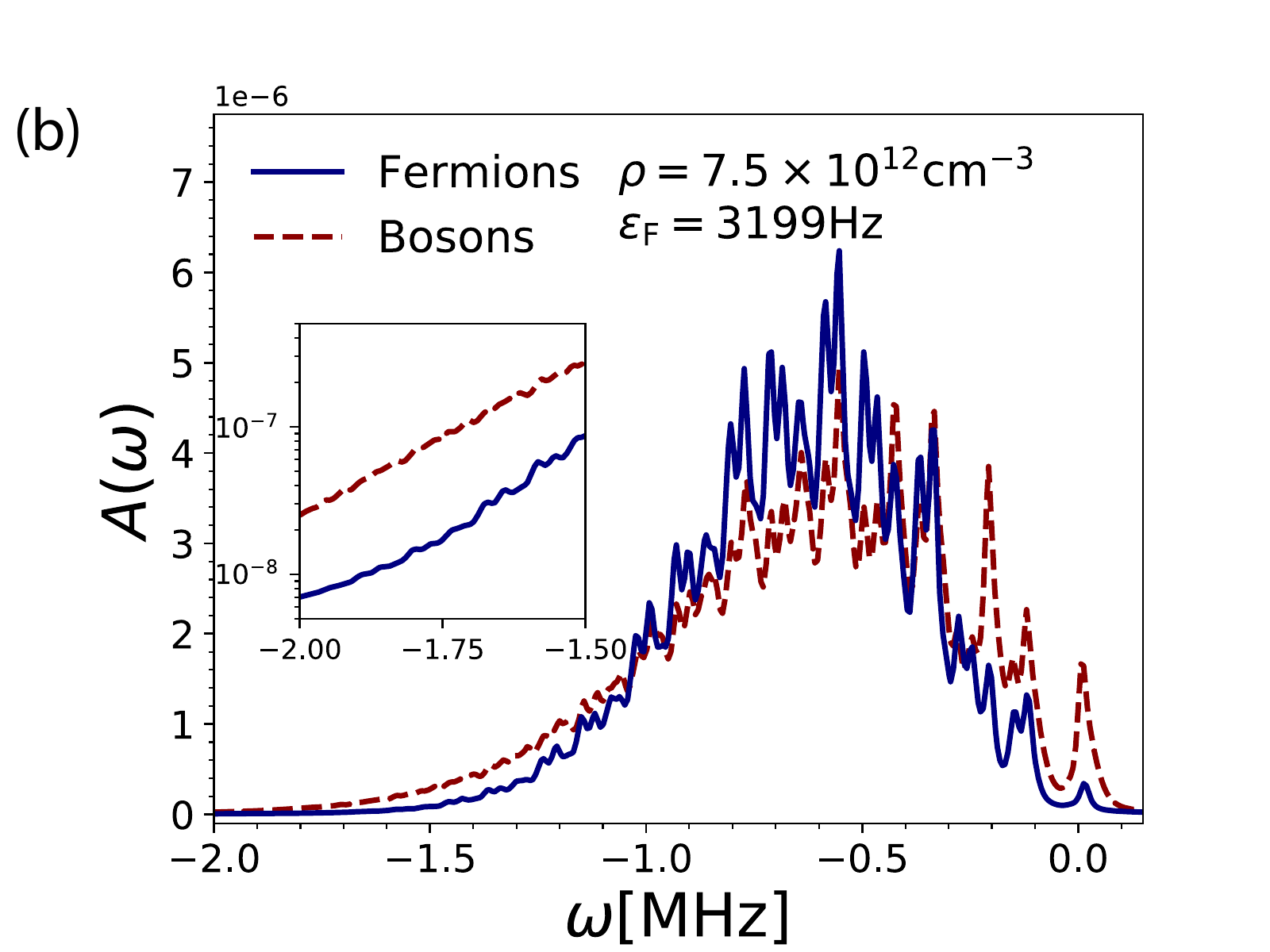}
\caption{\textbf{Shell structure with increasing densities.}  Comparison between spectra of a Rydberg impurity in a Fermi sea to that in a BEC at density (a) $\rho = 2.5\times10^{12}{\rm cm}^{-3}$ and (b) $\rho = 7.5\times 10^{12}{\rm cm}^{-3}$. In both cases, spectral suppression is visible for fermions at large detuning. At large densities an additional compression of the spectrum is found at small detunings that originates from suppressed density fluctuations in the fermionic medium. The response at large detuning [inset in (b)] reveals the formation of bound complexes of a large particle number. }
\label{fig4}
\end{figure}

We now turn to the effect of a larger density (red dashed line in Fig.~\ref{fig3}), corresponding to $\epsilon_{\rm F} \approx 3.2$KHz.  As an estimate, at this density, on average three atoms reside inside the Rydberg orbit, consistent with the spectral mean of $\sim 660$KHz found in our FDA calculation. The fact that the spectrum extends to larger detuning representing tetramers, pentamers, etc. is consistent with the Pauli-enforced Rydberg rotational blockade and the filling of the molecular shell structure.  At this density, the Fermi energy is high enough to populate up to $l=3$ states that have a significant spatial FCF with the region inside the Rydberg orbital. These states then participate in the dynamics and dress the impurity with molecular bound states. This is reflected in the corresponding filling of $s$, $p$,  $d$ and $f$ states in the shell structure, shown in the inset of Fig.~\ref{fig1}, which now includes a  contribution from $f$-shell states. As can be seen in Fig.~\ref{fignewRep} and Fig.~\ref{fig4}(b), in agreement with the relative occupation of shell states, spectral response is now  found up to frequencies that correspond to nine or more atoms bound to the impurity.  Note that the low-energy $l>0$ fermions deep below the Fermi surface are still unable to cross the barrier. Therefore, the number of particles that can participate in the dynamics will always be lower than that for bosons (at the same total number of atoms) leading to a suppression of weight at large detuning; see the discussion of Fig.~\ref{fig4} below.

As the gas density is increased, the spectrum evolves from resolved molecular lines with an asymmetric envelope (solid and dot-dashed lines in Fig.~\ref{fig3}) to a distribution of peaks that has a continuous envelope (dashed line), which moves progressively towards larger detuning. This broad spectral response represents the formation of Fermi superpolarons, the fermionic analog of superpolarons formed in Bose-Einstein condensates \cite{Rydbergpolaron1,Rydbergpolaron2,Rydbergpolaron3}. In Fig.~\ref{fig4}, we compare the spectrum of Fermi and Bose Rydberg superpolarons at large densities. As evident from Fig.~\ref{fig4}(a), compared to bosons,  Fermi statistics leads to a reduced spectral weight at large detuning due to the suppression of Franck-Condon overlaps for states of higher angular momentum $l$. As the density is, however, further increased,  a spectral suppression is observed not only at large but also at small detuning, see Fig.~\ref{fig4}(b).

This effect can be understood as follows: for ideal bosons, the positions of  particles are independent of each other and follow a Poisson distribution.  In contrast, quantum statistics imprints intrinsic density-density correlations onto fermions --- prominently visible as a correlation-hole at small particle separation --- that make them less compressible than bosons. The consequently reduced density fluctuations in the background gas then directly imply a suppressed spectral response both at large and small detuning. As an example, consider the case where a five-particle complex has the largest spectral weight.  In the Bose gas, fluctuations in the medium will strongly contribute to the spectral weight of four- and six-body complexes. But for fermions these contributions are weaker due to the suppressed density fluctuations, leading to a compression in the spectral response of fermions both at large and small detuning.  The observation of spectral compression can thus serve as a local probe of the compressibility of the many-body environment.

While the suppression at small detuning can qualitatively be understood in the simple picture of density fluctuations, capturing this effect quantitatively presents a challenge for theoretical approaches. In particular variational wave functions in terms of a perturbative expansion in particle-hole excitations \cite{Chevy2006,ChevyFermiPolaron} or vertex expansions in diagrammatic approaches \cite{Schmidt2010,Levinsen2013,Shi2018}, that were successfully applied in the description description of many conventional impurity problems,  are bound to fail to capture the physics of Rydberg impurities in a Fermi gas, due to their limitation in capturing the excitation of a large number of particle-hole excitations. In contrast the functional determinant approach can describe this system as it allows accounting for an arbitrary number of atoms transferred from the initial Fermi sea into the molecular shell structure, while keeping track of the full anti-symmetrization and thus the correct spatial nodal structure of the many-body wave function.

\section{Relation to Anderson orthogonality catastrophe} \label{OC}
The Anderson orthogonality catastrophe (OC) refers to the response of a Fermi gas to the introduction of a localized impurity \cite{AndersonOC}, where the generation of an infinity of particle-hole excitations leads to a new fermionic ground state that is orthogonal to the Fermi sea in absence of the impurity. This scenario was originally considered in the context of the X-ray absorption spectra in metals \cite{gogolin2004bosonization,Mahan}, where it is manifest in  characteristic threshold singularities and a power-law decay of the impurity Green's function in the time domain \cite{nozieres}. More recently,   signatures of the OC in the time domain have been shown to be accessible in Ramsey interference experiments of impurities in ultracold fermions controlled by Feshbach resonances \cite{RamseyOC2,RamseyOC1,fdg3}.

So far efforts to realize the physics of the OC in ultracold atoms \cite{RamseyOC2,RamseyOC1,fdg3,FermiPolaronRevRich}, following the observation of mobile Fermi \cite{fdg1,koschorreck_attractive_2012,RepulsivePolaronFermiExp,Zhang2012} and Bose polarons \cite{bec8,bec9,yan2019}, have  focused on systems where the impurity interacts with the Fermi gas via contact interactions that can support at most a single bound state. Realistic solid-state systems, such as quantum dots realized in semiconductor devices, go beyond this regime. In these systems, multiple electrons can be bound and the response of the Fermi environment of charge carriers can be probed through transitions between electronic states. These transitions lead to features characteristic of the OC, such as power-law edges in photoluminescence \cite{CardonaQD} and electron tunneling \cite{Geim,Khanin2005,Ruth}. 
While the Coulomb blockade limits the number of electrons that can occupy bound states in quantum dots, Rydberg impurities allow access to a regime where large multi-body bound complexes are formed in the presence of a fermionic environment.

In this work we focused specifically on the spectral signatures of molecular bound states. However, in our theoretical approach   the creation an infinite number of low-energy particle-hole excitations close to the Fermi surface  is also accounted for \cite{levitovFDA2,FermiPolaronRevRich} which  lead to asymmetric wings attached to each molecular peak. The existence of these molecular absorption edges opens a new avenue to study the OC in regimes  where the perturbing potential permits  multiple occupation of bound states. Unlike in  quantum dots,  the direct interaction between fermions can be tuned or entirely eliminated. Moreover, without direct interactions --- the case we consider in the present work --- the molecular peaks are inherently \textit{excited} states of the system, a scenario that goes beyond  limitations of previous studies in ultracold atomic and solid-state systems.

It is believed that in dimensions higher than one, a finite impurity recoil leads to the disappearance of the OC \cite{AchimKopp}.  An interesting question in the context of finite-mass Rydberg excitations is whether large impurity-bound molecules exhibit a crossover from quasiparticle to OC behavior.  The mass of larger complexes grows fast and should scale linearly with the number of atoms bound to the impurity. Thus larger complexes may show a sharp OC behavior, while smaller ones may exhibit quasiparticle properties.    This  argument also implies that our approach to infinitely heavy impurities may in fact be a reasonable starting point to describe large molecular polarons even in the regime where the impurity mass is finite. This unusual behavior makes Rydberg excitations a unique setting for exploring OC physics, which so far has been challenging in Feshbach-coupled quantum gases for which recoil effects are significant \cite{fdg3}.

Furthermore, since the interaction range varies with the principal number $n$ as $r_0\sim n^2 \hbar^2 4 \pi \epsilon_0/e^2 m$, the energy and the size of the molecular states can be tuned to the different regimes of dynamics. Along with control over temperature and gas density, this opens a door to many compelling questions, such as whether the exchange of medium fluctuations may lead to a hybridization of bound states and whether the OC has an observable effect on the hybridization dynamics.

\section{Conclusion} \label{Conc}

We describe a platform to study multiscale quantum impurity effects in a Fermi sea. To this end, we have developed a time-dependent many-body formalism based on functional determinants that treats the bound Rydberg molecules and scattering states on the same footing, to follow the non-equilibrium time evolution of the degenerate Fermi gas interacting with a spatially extended Rydberg impurity.  Accounting for all angular momentum  contributions,  we predict the absorption spectrum and demonstrate that the spectroscopy of Rydberg excitations allows probing the physics of quantum impurities interacting with a Fermi gas in a regime where the impurity itself may extend over to nearly the system size.

The analysis of the absorption spectra reveals the emergent effect of a Pauli-enforced Rydberg rotational blockade that inhibits the formation of  trimers, tetramers, and higher-order oligomer molecules.  While blockade effects are at play in the formation of fermionic ultracold Feshbach molecules \cite{DeMarco2019}, there, the rotational barriers are at much higher energies. In contrast, by virtue of the fact that Rydberg molecules are created at enormous distances, their rotational barriers are but a small fraction of the binding energies so that the physics of rotational blockade can be studied in a previously inaccessible regime.

At low  densities of the environment, the absorption response reveals a new shell structure where fermions subsequently fill shells that are defined by angular momentum and vibrational quantum numbers. In the many-body regime at large densities our findings deviate from the expectation that, for such macroscopic fermion systems, the spectral response should resemble that of a Rydberg impurity in a Bose gas. Instead we find that the macroscopic occupation of angular momentum shell modes results in a Fermi superpolaronic response whose signature is a spectral narrowing (Fermi compression) compared to the response in a bosonic gas, which can  serve as a local probe of the density-density correlations of the many-body environment, complementing probes of transient dynamics in quantum gases \cite{ZwierleinPrivComm}.

The evolution from few- to many-body behavior involves bound-state physics in the presence of an environment  that goes beyond condensed matter realizations, such as phonon-induced binding of Cooper pairs \cite{CooperPair} or trion formation in presence of a charge-carrier environment in doped two-dimensional semiconductors \cite{Atac1,Reichman2019}. In particular, the exquisite control over Rydberg excitation represents a novel knob for structuring the local density surrounding the Rydberg atoms and permits  study of  the interplay between non-perturbative, few- and many-body  effects in Fermi systems.

\noindent\textbf{\textit{Experimental realization.---}} To realize the physics of Fermi superpolarons and study the suppression of molecule formation, experiments with mass-imbalanced Bose-Fermi or Fermi-Fermi mixtures featuring a low density of a heavy atomic species in a background Fermi gas of lighter atoms are ideal. Starting from impurities initially in their atomic ground state, a laser transition implements the sudden quench of the impurity potential seen by the atoms in the Fermi sea that are initially uncorrelated to the impurity atom. In order to avoid blockade effects and to gain access  to the linear response regime one can use an excitation pulse of sufficiently low power~\cite{Rydbergpolaron2}.

Alternatively, one may also employ Rydberg excitations in a single-component Fermi gas. Indeed, first evidence of the Fermi suppression discussed in this work have recently been observed in a non-degenerate gas of $^{87}$Sr \cite{whalen2019}. In such a setup, a laser excites a fermion from the single-component Fermi gas to a highly excited Rydberg state. Since the initial state of the Fermi gas corresponds to a Slater determinant, it exhibits density-density correlations with a characteristic correlation hole at short distances. This correlation hole leaves its trace in a suppression of the formation rate of dimers, as the Rydberg excitation is created from the Fermi gas itself. While this effect significantly influences the absorption signal of dimers relative to the scenario of a two-component gas studied in this work,  in the limit of large densities the two scenarios will not differ significantly since  the initial correlations of the Rydberg impurity and the bath become increasingly irrelevant. In consequence,  single-component quantum gases will also realize the physics of Fermi superpolarons and suppressed molecular formation discussed in this work.

We note that in the present work we have focused on the zero temperature response. Our results extend, however, to finite temperatures less than the Fermi temperature, {\em i.e.} in the range $\sim25$-$150$nK for the densities studied in this work. This temperature range can be further increased by employing a fermionic species of a small  mass such as atomic $^6$Li.

\noindent\textbf{\textit{Outlook.---}} The presented results are exact for a localized impurity and approximate for heavy, but mobile impurities. For infinitely heavy impurities, the orthogonality catastrophe governs the many-body dynamics and quasiparticles are absent due to the generation of an infinite number of particle-hole excitations. For a finite-mass impurity, the energy cost of impurity recoil suppresses these low-energy fluctuations leading to a finite quasiparticle weight of Fermi polarons \cite{ProkofevKagan,AchimKopp}. Moreover, for impurities interacting with the environment via short-range interactions, impurity recoil leads to a transition between a polaron and a molecular ground state \cite{Punk2009,Bruun2010,Schmidt2011}. In the case of Rydberg impurities, the consequences of a finite impurity mass are currently unknown. In which way Fermi polarons form, and whether polaron-to-molecule transitions can occur are  open questions.  Additionally, since the multi-body bound molecules are inherently excited states of the many-body system, the question arises whether impurity motion can potentially lead to recoil-induced decay between the different molecular states, rendering them unstable. While these effects can be experimentally addressed by exploiting the long coherence times accessible with current technologies, their theoretical investigation requires new many-body approaches that can capture the multiscale nature of Rydberg impurities as well as their motion.

\begin{acknowledgments}
We thank M.~Wagner for helpful discussions. J.~S. acknowledges support from a visiting student fellowship at the Institute for Theoretical Atomic, Molecular, and Optical Physics (ITAMP) at Harvard University and the Smithsonian Astrophysical Observatory, and the Natural Sciences and Engineering Research Council of Canada (NSERC), and  the hospitality of the Department of Physics at Harvard University, the Stewart Blusson Quantum Matter Institute at the University of British Columbia and the Department of Physics at the Technical University of Munich. H.~R.~S. acknowledges support from the National Science Foundation (NSF) through a grant for ITAMP at Harvard University and the Smithsonian Astrophysical Observatory. T.~C.~K. acknowledges support from AFOSR (FA9550-14-1-0007) and the Robert A. Welch Foundation (C-1844). E.~D. acknowledges support from the Harvard-MIT CUA, AFOSR-MURI: Quantum Phases of Matter (grant FA9550-14-1-0035) and AFOSR-MURI: Photonic Quantum Matter (award FA95501610323). R.~S. is supported by the Deutsche Forschungsgemeinschaft (DFG, German Research Foundation) under Germany's Excellence Strategy -- EXC-2111 -- 390814868. 
\end{acknowledgments}

\appendix

\section{Deriving $A(\omega)$ from $S(t)$} \label{AbsS}
In linear response theory, the two-photon absorption spectrum is given by Fermi's golden rule
\begin{equation}\label{eqn::Aws}
A(\omega) = 2\pi \sum_{i,f} \frac{e^{-\beta(E_i - \mu n_i)}}{Z} |\bra{f} \hat \Omega_L  \ket{i}|^2 \delta\big(\omega - (E_f - E_i)\big),
\end{equation}
where $\beta^{-1} = k_B T$, $Z$ is the grand canonical partition function, $\mu$ is the chemical potential, and $n_i$ denotes the number of particles present in the initial state. As before, we have set $\hbar =1$.  In the experiment, $\hat \Omega_L$ represents the laser-induced transition operator $\ket{\uparrow}\bra{\downarrow}$ that excites an atom to a Rydberg state $\ket{\uparrow}$, which, in our model, takes the form $\hat \Omega_L = \hat{d}^\dagger$ for the impurity in the atomic gas. The sum in equation (\ref{eqn::Aws}) extends over complete sets of initial $\ket{i}$ and final $\ket{f}$ many-body states with energies $E_i$ and $E_f$.

Inserting the Fourier representation of the delta distribution and using that $\ket{i}$ and $\ket{f}$ are many-body eigenstates of $\hat{H}_0$ and $\hat{H}$, respectively, $A(\omega)$ takes the form
\begin{eqnarray} 
A(\omega) &=& \int^{\infty}_{-\infty} {dt}  \sum_{i,f} \bra{i}  \hat{\mathcal{\varrho}} e^{i \hat{H}_0 t}\ket{f} \bra{f}e^{-i \hat{H} t} \ket{i} e^{i\omega t} \nonumber \\
&=& \int^{\infty}_{-\infty} {dt} {\rm Tr} \left[ \hat{\mathcal{\varrho}} e^{i \hat{H}_0 t}e^{-i \hat{H} t} \right]e^{i\omega t} \nonumber \\
&=& 2 {\rm Re} \int^{\infty}_{0} {dt}  {\rm Tr} \left[ \hat{\mathcal{\varrho}} e^{i \hat{H}_0 t}e^{-i \hat{H} t} \right]e^{i\omega t},
\end{eqnarray}
where $\hat{\mathcal{\varrho}}$ is the density matrix of the initial state of the non-interacting particles.
The time-dependent overlap is
\begin{eqnarray} 
S(t) = {\rm Tr} [e^{i \hat{H}_0 t}e^{-i \hat{H} t} \hat{\varrho} ]
\end{eqnarray}
and we arrive at the relation \cite{FermiPolaronRevRich}
\begin{eqnarray} 
A(\omega) = 2 {\rm Re} \int_0^\infty dt e^{i\omega t} S(t),
\end{eqnarray}
which is Eq. (\ref{eqn::AS}) of the main text.

\section{Functional determinant approach (FDA)}\label{FDA_Appendix}
The functional determinant approach (FDA) provides an effective evaluation of many-body expectation values in Fock space in terms of first quantized single-particle operators \cite{levitovFDA1}. For bilinear operators, one finds the relation \cite{KlichFDA}
\begin{eqnarray} \label{FDAFormula}
\langle e^{\hat{X}_N} \cdots e^{\hat{X}_1}\rangle_T = {\rm det} [1 - \zeta \hat{n} + \zeta e^{\hat{x}_N} \cdots e^{\hat{x}_1} \hat{n} ]^{\zeta}.
\end{eqnarray}
$\langle \quad \rangle_T$ denotes the thermal expectation value evaluated with respect to the appropriate thermal density matrix, $\hat{X}_i$ denotes a quadratic (bilinear) many-body operator and $\hat{x}_i$ is the corresponding single-particle operator. More specifically, $\hat{X} = \sum_{j,k}  \bra{j}\hat{x}\ket{k} \hat{a}^{\dagger}_j \hat{a}_k$, where $\hat{a}_i$ are second-quantized operators defined with respect to the single-particle basis states $\ket{i}$, $\hat{n}$ is the single-particle number operator and $\zeta = 1$ for fermions and $\zeta = -1$ for bosons. For more details about the FDA for fermions we refer to Refs.~\cite{levitovFDA1,levitovFDA2,KlichFDA,FDA2} and for bosons to Refs.~\cite{KlichFDA,Rydbergpolaron1}.

\section{Time-dependent overlap of a Rydberg impurity in a Fermi sea $S(t)$} \label{SPS}
For the Hamiltonian Eq. (\ref{eqn::Ham}), the FDA formula (\ref{FDAFormula}) gives
\begin{equation}
\langle e^{i \hat{H}_0 t}e^{-i \hat{H} t} \rangle_T = {\rm{det}}[1 - \hat{n}_{{\rm FD}} + \hat{n}_{{\rm FD}} e^{i\hat{h}_0 t}e^{-i\hat{h}t}],
\end{equation}
which yields Eq. (\ref{FDA}) in the main text
\begin{equation}
S(t) = {\rm{det}}[1 - \hat{n}_{{\rm FD}} + \hat{n}_{{\rm FD}} e^{i\hat{h}_0 t}e^{-i\hat{h}t}].
\end{equation}

To find $S(t)$, we compute the matrix elements of the temperature($T$)-dependent single-particle operator 
\begin{equation}\label{FDA_C_f}
\hat{C}(T)= 1 - \hat{n}_{{\rm FD}} + \hat{n}_{{\rm FD}} e^{i\hat{h}_0 t}e^{-i\hat{h}t}
\end{equation}
in the basis of non-interacting single-particle orbitals. We insert in Eq. (\ref{FDA_C_f}) the complete set of eigenstates of the interacting single-particle Hamiltonian $\hat{h}$, determined from the solutions of the Schr{\"o}dinger equation
\begin{equation} \label{F_SE}
[-\frac{1}{2m}\nabla^2+V_{\rm{Ryd}}({\bf r})] \phi({\bf r})= E\phi({\bf r}),
\end{equation}
where $m$ is the mass of bath atoms.  For simplicity, and to allow direct comparison with boson spectra, we use the mass of $^{87}$Rb in our FDA calculations for both fermions and bosons.

The spherically symmetric Rydberg potential leads to separation of angular and radial variables: ${\phi({\bf r})} = \frac{u_{kl}(r)}{r} Y_{lm}(\bf{\Omega})$, reducing the problem to the radial Schr{\"o}dinger equation
\begin{equation} \label{RSE}
[-\frac{1}{2m}\frac{\partial^2}{\partial r^2}+ \frac{l(l+1)}{2mr^2}+V_{{\rm{Ryd}}}(r)] u_{kl}(r)= \epsilon_{kl} u_{kl}(r),
\end{equation}
which we solve numerically in a spherical box of radius $R =2\times 10^5 a_0$. We provide  details about the numerical diagonalization in Appendix \ref{Numerics}.

Expressed in the non-interacting single-particle states $\ket{s}$ with energy $\epsilon_{s}$ [$s = (k_{s},l_{s},m_{s})$ with nodal number $k$, angular momentum $l$ of projection $m$ on the quantization-axis], one finds
\begin{eqnarray}\label{FDA_C_f_basis}
C_{s',s} (T) &\equiv& \bra{s'} \hat{C} \ket{s} \nonumber \\
& = & \delta_{s,s'} \Big(1- n_{\rm FD}(\epsilon_{s})\Big) + n_{\rm FD}(\epsilon_{s'}) \bra{s'} e^{i\hat{h}_0 t}e^{-i\hat{h}t} \ket{s}. \nonumber \\
\end{eqnarray}
At $T=0$, fermions fill up single-particle states $\epsilon_{s} \leq \epsilon_{\rm F}$ and $n(\epsilon_{s}) = 1$. Eq. (\ref{FDA_C_f_basis}) simplifies to
\begin{eqnarray}\label{FDA_C_f_basis_0}
C_{s',s} (0)=\begin{cases}
    \bra{s'} e^{i\hat{h}_0 t}e^{-i\hat{h}t} \ket{s}, & \text{if $\epsilon_s < \epsilon_{\rm F}$}.\\
    \delta_{s,s'}, & \text{otherwise}.
  \end{cases}
\end{eqnarray}

We insert a complete set of the `interacting' single-particle states $\ket{\alpha}$, where $\alpha = (k_{\alpha},l_{\alpha},m_{\alpha})$ labels the interacting single-particle states with energy $\omega_{\alpha}$,
\begin{eqnarray}
C_{s',s} (0) &=&  \sum_{\alpha} \bra{s'} e^{i\hat{h}_0 t}e^{-i\hat{h}t} \ket{\alpha}\bra{\alpha}\ket{s} \nonumber \\
&=& e^{i\epsilon_{s'} t} \sum_{{\alpha}} e^{-i\omega_{\alpha}t} \bra{s'}\ket{{\alpha}}\bra{{\alpha}}\ket{s}.
\end{eqnarray}
Due to spherical symmetry the impurity potential does not couple different angular momentum states: $\bra{{\alpha}}\ket{s} \equiv \bra{k_{\alpha},l_{\alpha},m_{\alpha}}\ket{k_{s},l_{s},m_{s}} =  \bra{k_\alpha}\ket{k_s} \delta_{l_{\alpha},l_s}\delta_{m_{\alpha},m_s} $. Thus
\begin{eqnarray}
C_{s',s} (0)  &=& \delta_{l_{s},l_{s'}} \delta_{m_{s},m_{s'}} \sum_{k_{\alpha}} e^{i(\epsilon_{s'}-\omega_{{\alpha}})t}\bra{k_{s'}}\ket{k_{\alpha}} \bra{k_{\alpha}}\ket{k_s} \nonumber \\
&=& \delta_{l_{s},l_{s'}} \delta_{m_{s},m_{s'}} \mathcal{C}_{s',s}(l,m),
\end{eqnarray}
where we have defined the matrix ${\bf \mathcal{C}}_{s',s}(l,m) =  \sum_{k_{\alpha}} e^{i(\epsilon_{s'}-\omega_{{\alpha}})t}\bra{k_{s'}}\ket{k_{\alpha}} \bra{k_{\alpha}}\ket{k_s}$ that couples different nodal $k_s$ states within the subspace labeled by $l_s$ and $m_s$. 
In matrix notation one thus finds a block diagonal structure: ${\bf \mathcal{C}} = {\rm diag}\{{\bf \mathcal{C}}(l = 0,m=0) {\bf \mathcal{C}}(l = 1,m=-1) {\bf \mathcal{C}}(l=1,m=0) {\bf \mathcal{C}}(l=1,m=1) \ldots {\bf \mathcal{C}}(l=l_{{\rm max}},m=m_{{\rm max}}) \}$, where $l_{\rm max}$ and $m_{\rm max}$ label the highest occupied states at or just below the Fermi level.
The determinant of a block diagonal matrix is the product of determinants of the diagonal blocks
\begin{eqnarray}
S(t) = \prod_{l,m} s_{l,m}(t),
\end{eqnarray}
where $s_{l,m}(t) = {\rm det} \left[{\bf \mathcal{C}}(l,m)\right]$. Accounting for the degenerate $m$ states within the manifold of $l$ leads to
\begin{eqnarray}
S(t) = \prod_{l}  [s_{l} (t)]^{2l+1},
\end{eqnarray}
with $s_{l}(t) = {\rm det} {\bf \mathcal{C}}(l) =  \sum_{k_{\alpha}} e^{i(\epsilon_{k_s',l_s'}-\omega_{k_{\alpha},l_{\alpha}})t}\bra{k_{s'}}\ket{k_{\alpha}} \bra{k_{\alpha}}\ket{k_s}$ that depends only on $l$. We defined $S^{l}(t) = [s_{l} (t)]^{2l+1}$ in the main text.

\section{Time-dependent overlap of a Rydberg impurity in a Bose-Einstein condensate (BEC)}\label{StBEC}
The Fock state representing the macroscopic occupation of bosons in an ideal BEC is given by $\ket{\Psi_{\rm BEC}} = \frac{1}{\sqrt{N_{\rm B}!}} (\hat{b}_0^{\dagger})^{N_{\rm B}} \ket{{\rm vac}}$, where $\hat{b}_0^\dagger$ is the creation operator of a boson in the lowest-energy single-particle state of zero angular momentum, $N_{\rm B}$ is the boson particle number and $\ket{{\rm vac}}$ is the vacuum state. Evaluating the time-dependent overlap Eq.~\eqref{eqn::LE} with respect to the density matrix $\hat{\varrho}_{\rm BEC} = \ket{\Psi_{\rm BEC}} \bra{\Psi_{\rm BEC}}$ gives
\begin{eqnarray}
S_{\rm BEC}(t) = \bra{\Psi_{\rm BEC}} e^{i \hat{H}_0 t}e^{-i \hat{H} t } \ket{\Psi_{\rm BEC}}.
\end{eqnarray}
Expanding $\ket{\Psi_{\rm BEC}}$, we find
\begin{eqnarray}
S_{\rm {BEC}}(t) & =  & e^{i N_{\rm B} \epsilon_0 t} \frac{1}{N_{\rm B}!} \nonumber \\ 
\times & \bra{{\rm vac}} & \Big(\sum_{\alpha } \bra{0}\ket{\alpha} \hat{b}_{\alpha}\Big)^{N_{\rm B}} e^{-i \hat{H} t }   \Big(\sum_{\alpha } \bra{\alpha}\ket{0} \hat{b}^{\dagger}_{\alpha}\Big)^{N_{\rm B}} \ket{{\rm vac}} \nonumber \\
&=&  \Big(\sum_{\alpha, \alpha' } e^{i (\epsilon_0 - \omega_{\alpha'}) t}  \bra{0}\ket{\alpha} \bra{\alpha'}\ket{0}    \delta_{\alpha,\alpha'} \Big)^{N_{\rm B}}\nonumber \\
&=&  \Big(\sum_{\alpha}  \left|\bra{\alpha}\ket{0}\right|^2 e^{i (\epsilon_0 - \omega_{\alpha}) t}  \Big)^{N_{\rm B}}. \label{SBEC}
\end{eqnarray}

\vspace{-4mm}
\subsection{BEC peak densitiy $\rho_0$ used in comparison of the spectrum of the Rydberg impurity in a Fermi sea to that in a BEC} \label{PeakBEC}
To compute Fermi and Bose Rydberg polaron responses, we must ensure that the same densities at the position of the impurity are used in the calculation. To this end, for a given density of fermions in the Fermi gas $\rho$, we determine $N_{\rm B}$ that corresponds to the peak density in the BEC $\rho_0$ at the center of the trap given by
\begin{eqnarray}
\rho_0 &=& N_{\rm B}\left|\Psi_{\rm BEC}\right|^2 \nonumber \\
&=& N_{\rm B} \left| \frac{1}{2} \sqrt{\frac{1}{\pi}} \sqrt{\frac{2}{L}} \frac{{\rm sin}(\pi r /L)}{r}\right|^2_{r\rightarrow 0} \nonumber \\
&=& N_{\rm B} \frac{\pi}{2L^3},
\end{eqnarray}
where we quantized the wave function in a spherical box of radial extent $L$. We then use $N_{\rm B} = \rho_0 \frac{2L^3}{\pi}$ in $S_{\rm BEC}(t)$, Eq. (\ref{SBEC}), and choose $\rho=\rho_0$ for the corresponding fermionic system.

\vspace{-4mm}
\section{Numerical procedures}\label{Numerics}
\subsection{Exact diagonalization of the radial Schr{\"o}dinger equation}
We solve the radial Schr{\"o}dinger equation (\ref{RSE}) numerically in a spherical box imposing hard-wall boundary conditions. The Hamiltonian matrix is constructed in the basis states $\ket{r}$ that serve as discrete real-space representation of states in the continuous radial coordinate $r$. 

The numerical solutions determine the radial wave functions $u_{k_{\alpha}}(r) = \bra{r}\ket{k_{\alpha}}$ for the interacting states and $u_{k_{s}}(r) = \bra{r}\ket{k_{s}}$ for the non-interacting states. The discretization scheme defines the integration measure $\int dr = \sum_i \Delta_{r_i}$, where $\Delta_{r_i}$ is the grid-step size at point $r_i$. Using this integration measure, we construct normalized wavefuctions.

\subsection{Construction of overlap matrix $\mathcal C$}
We construct the matrix $C_{s',s} (0)$ in Eq. (\ref{FDA_C_f_basis_0}) in the single-particle basis that constitutes the spin-polarized Fermi sea $\ket{\Psi_{\rm FS}}$. In this construction, the Fermi energy $\epsilon_{\rm F} = \frac{ (6 \pi^2 \rho)^{2/3}}{2 m}$  sets the number $k_{s_{\rm max}}^{(l)}$ which determines the uppermost single-particle state of angular momentum $l$ filled by the Fermi sea.

\vspace{-4mm}
\subsection{Numerical computation of the time-dependent overlap $S(t)$}
In computing time-dependent overlaps $S(t)$ we insert a complete set of interacting states $\ket{k_{\alpha}}$, including up to $2 \times k_{s_{\rm max}}^{(l)}$ states for each angular momentum sector $l$. To eliminate remaining truncation errors, we normalize $s_l(t)$ according to $s_l(t) \rightarrow s_l(t)/s_l(0)$ before taking the power to $2l+1$.

\vspace{-4mm}
\subsection{Fourier transform of $S(t)$}
To stabilize $A(\omega)$ against the Gibbs phenomenon when performing the numerical Fourier transform of $S(t)$, we impose an exponential decay on the computed $S(t)$, {\em i.e.} $S(t) \rightarrow S(t) e^{-t/t_{\eta}}$. This procedure results in absorption peaks in $A(\omega)$  with finite width $\sim 1/t_\eta$.

\section{Analytical dependence of Franck-Condon overlaps on the angular momentum $l$}\label{Analytic}
We consider Franck-Condon overlaps between the lowest bound dimer state $\ket{1_{\alpha},l}$ and free states $\ket{k_s,l}$ of nodal number $k_s$: 
\begin{eqnarray}
{\cal I}  &=& \bra{1_{\alpha},l}\ket{k_s,l} = \int dr \bra{1_{\alpha},l}\ket{r}\bra{r}\ket{k_s,l}  \nonumber \\
&\equiv&  \int dr ~{u_{1_{\alpha},l}^*(r)} ~ u_{k_s,l} (r) \nonumber \\
&\propto& \int dr~ j_{l}({\mathcal K}^{(l)}_{k_s} r) \delta(r-R_0) \nonumber \\
&=& j_{l}({\mathcal K}^{(l)}_{k_s} R_0),
\end{eqnarray}
where $j_l(x)$ is the spherical Bessel function of order $l$ and argument $x$. We have inserted a resolution of the identity $\int dr \ket{r}\bra{r}\equiv 1$ in the first line, used the definitions of the interacting and non-interacting wave functions  in the second line, and in the third line we have approximated, up to a constant, the lowest bound state as a delta function. This is a reasonable approximation to derive the dependence on $l$, as the dimer wave function is predominantly bound to the approximately harmonic well at $R_0$, see Figs.~\ref{fig2} and \ref{fig6}(b).

The free solutions satisfy the boundary condition $j_{l}({\mathcal K}^{(l)}_{k_s} R) \approx \frac{\sin({\mathcal K}^{(l)}_{k_s} R - l \pi/2)}{{\mathcal K}^{(l)}_{k_s} R}= 0$ as $R\rightarrow\infty$ and for sufficiently large $l$.  From this follows the solution ${\mathcal K}^{(l)}_{k_s}= \frac{(n+l/2)\pi}{R}$ with $n \in \mathbb{N}$. We thus find 
\begin{eqnarray}
{\cal I}  &\sim& j_{l}\big({\pi(n+l/2)}\frac{R_0}{R} \big).
\end{eqnarray}
Since $R_0\ll R$, we can derive the dependence on $l$ by Taylor expanding ${\cal I}$ for small arguments ${\pi(n+l/2)}\frac{R_0}{R} \ll 1$.  Expressing the spherical Bessel functions $j_{l}(x)$ as Bessel functions of the first kind, which can be readily expanded as a series, we find
\begin{eqnarray}
{\cal I}  &\sim& \frac{\sqrt{\pi}(\pi R_0/R)^l ~(n+l/2)^l ~2^{-1-l}}{\Gamma(l+3/2)}.
\end{eqnarray}
Further, we simplify the Gamma function using Sterling's formula to obtain
\begin{eqnarray}
{\cal I}  &\sim&  \frac{1}{\sqrt{2}} \Big(\pi \frac{R_0}{R}\Big)^l ~(n+l/2)^l ~\sqrt{l+3/2} \nonumber \\
&&\quad\quad\quad\quad\quad \times ~2^{-1-l} ~\Big(\frac{l+3/2}{e}\Big)^{-l-3/2},
\end{eqnarray}
which shows the leading super-exponential dependence on $l$ discussed in the main text.


%

  \end{document}